\documentclass[prb,preprint]{revtex4-1}
\usepackage{bm}
\usepackage{stmaryrd}
\usepackage{amsmath}
\usepackage{amssymb}
\usepackage{graphicx}
\usepackage{graphics}
\usepackage{textcomp}
\usepackage{calrsfs}
\usepackage{yfonts}
\usepackage{amsthm}
\usepackage{lipsum}
\usepackage{amsfonts}
\newcommand{\p}{{\partial}}

\newcommand{\la}{{\langle}}
\newcommand{\ra}{{\rangle}}

\newcommand{\haf}{{\frac{1}{2}}}

\newcommand{\be}{\begin{equation}}
\newcommand{\ee}{\end{equation}}
\newcommand{\bea}{\begin{eqnarray}}
\newcommand{\eea}{\end{eqnarray}}
\begin{document}
\title{Quantum propagator and characteristic equation in the presence of a chain of $\delta$-potentials}

\author{A. Refaei}
\email{refaei@iausdj.ac.ir}
\affiliation{Department of Physics, Sanandaj branch, Islamic Azad University, Sanandaj, Iran.}

\author{F. Kheirandish}
\email{fkheirandish@yahoo.com}
\affiliation{Department of Physics,
Faculty of Science, University of Isfahan,
Isfahan, Iran.}

\date{\today}
\begin{abstract}
The quantum propagator and characteristic equation in the presence of a chain of $\delta$-potentials are obtained in the rectangular, cylindrical and spherical coordinate systems. The simplicity and efficiency of the method is illustrated via examples. As an application, the characteristic equation of a quantum harmonic oscillator confined to an infinite box is obtained. The roots of the characteristic equation, determining the energy eigenvalues of the restricted oscillator, are calculated approximately and compared with the existing numerical data.
\end{abstract}
\keywords{Quantum propagator, Characteristic equation, $\delta$-potential; Green's function; Spectrum}
\maketitle

\section{Introduction}

An important two-point function in investigation of the dynamics of an arbitrary quantum system is the quantum propagator or kernel. Knowing the quantum propagator, the exact time-evolution of the system can be obtained from the initial state. Among the interesting problems in both theoretical and applied quantum mechanics are the investigation of the confined systems or quantum dynamics in the presence of some constraints usually of Dirichlet type. In some situations, by choosing a suitable coordinate system, the constraints can be modelled by a linear combination of the $\delta$-potentials and then the problem is to find the quantum propagator in the presence of a chain of the $\delta$-potentials. The coefficients of this expansion are coupling constants which in the presence of impenetrable walls or equivalently strong coupling limit, are set to infinity at the end of the calculations. The quantum propagator or kernel is in fact the Green's function of the Schr\"{o}dinger equation \cite{Sakurai}.

Green's functions originally invented to provide solutions to electrostatic problems \cite{Green}, and subsequently its techniques has grown to encompass an immense number of disparate subjects, specifically in quantum field theory, electrodynamics and statistical field theory \cite{Appl}. The Green's function methods of the quantum field theory have become generally recognized as a powerful mathematical tool for studying the complex interacting systems. The analytic properties of the Green's function in the complex energy plane accounts for important physical properties of the system. Here we need to determine the Green's functions in the presence of the $\delta$-potentials.

The $\delta$-potentials have been studied using a variety of techniques, see for example \cite{4,5,6}. The $\delta$-potentials have many interesting applications, for example in Casimir energy calculations \cite{Milton2}, confinement problems \cite{confi}, path integral methods \cite{Grosche}, attractive $\delta$-potential Bose gas \cite{Tsubasa}, periodic integrable systems \cite{Erdal} and atom-Laser interaction \cite{Campo}. In confinement problems, we can assume the boundaries as impenetrable walls $(\lambda\rightarrow\infty)$, in this case the energy eigenvalues of the constraint system are determined as the poles of the frequency-space Green's function or equivalently as the roots of the characteristic equation. Among the interesting constrained problems is the quantum harmonic oscillator confined to an infinite one-dimensional box. This problem has been investigated by many authors see for example \cite{Campoy}, \cite{Aguilera}, and references therein. In most of those investigations the problem has been treated numerically. Here a closed form expression for the characteristic equation of the constrained quantum harmonic oscillator is introduced.

The outline of this article is as follows. In Sec. II and its subsections, the quantum propagator, as the Green's function of the Schr\"{o}dinger equation, is investigated in the rectangular, cylindrical and spherical coordinate systems based on the symmetric considerations. In each coordinate system, the reduced Green's function is obtained in the presence of a chain of $\delta$-potentials. The solution is based on generalizations of the method introduced in \cite{main} where the Green's function is obtained in the presence of a single $\delta$-potential. We will see that in order to find the reduced Green's function we only need to find the inverse of an algebraic matrix defined by the values of the free-space reduced Green's function at the boundary points. The efficiency and simplicity of the method is illustrated via examples. In Sec. III, as an applications of the method, the characteristic equation of a quantum harmonic oscillator confined to an infinite-box is obtained. The roots of the characteristic equation, determining the energy eigenvalues, can be calculated easily to any order of approximation. The first six energy eigenvalues of the constrained oscillator are calculated approximately and compared with the numerical data reported in \cite{Aguilera}. Concluding remarks are given in Sec. IV and finally in Sec. V a suggested problem is given.

\section{Quantum Propagator}

In a general coordinate system the quantum propagator satisfies \cite{Sakurai}
\begin{equation}\label{P1}
 \bigg[ -\frac{\hbar^2}{2m}\nabla^2+V(\mathbf{x})-i\hbar\frac{\p}{\p t}\bigg]K(\mathbf{x},t;\mathbf{x}',t')=-i\hbar\delta(\mathbf{x}-\mathbf{x}')
 \delta(t-t'),
\end{equation}
showing that the propagator is in fact the Green's function of the Schr\"{o}dinger equation. By Inserting the Fourier transform
\begin{equation}\label{F}
 K(\mathbf{x},t;\mathbf{x}',t')=\int \frac{d\omega}{2\pi}\,e^{-i\omega (t-t')}\tilde{K}(\mathbf{x},\mathbf{x}',\omega),
\end{equation}
into Eq. (\ref{P1}) and multiplying across by $-2m/\hbar^2$, we find
\begin{equation}\label{FS}
 \bigg[\nabla^2-\frac{2m}{\hbar^2}\,V(\mathbf{x})+\frac{2m\omega}{\hbar}\bigg]\tilde{K}(\mathbf{x},\mathbf{x}',\omega)=
 \frac{2mi}{\hbar}\delta(\mathbf{x}-\mathbf{x}').
\end{equation}
 In the following sections we solve this equation in the rectangular, cylindrical and spherical coordinate systems when the potential $V(\mathbf{x})$ is a linear combination of the $\delta$-potentials.

\subsection{Rectangular Geometry}

In the rectangular geometry we assume that the walls are located at the positions $z=a_1,\,z=a_2,\,\cdots,z=a_n$, and the potential is defined by
$V(z)=\sum\limits_{j=1}^n \mu_j \,\delta(z-a_j)$, where $\mu_j$ are coupling constants which in the case of strong coupling limit or impenetrable walls tend to infinity. Due to the transverse symmetry, the frequency-space propagator can be Fourier transformed as
 \begin{equation}\label{SFR}
   \tilde{K}(\mathbf{x},\mathbf{x}',\omega)=\int\int\frac{dk_x}{2\pi}\frac{dk_y}{2\pi}\,e^{ik_x (x-x')}e^{ik_y (y-y')}G^{(r)}(z,z';\omega,k_x,k_y),
 \end{equation}
now by inserting Eq. (\ref{SFR}) into Eq. (\ref{FS}) we find an equation for the reduced Green's function as
\begin{equation}\label{RGF}
  \bigg[\frac{d^2}{d z^2}-k_0^2-\sum\limits_{j=1}^n \lambda_j \,\delta(z-a_j)\bigg]\,g^{(r)}(z,z';\omega,k_x,k_y)=-\delta(z-z'),
\end{equation}
where
\begin{eqnarray}
  g^{(r)}(z,z';\omega,k_x,k_y) &=& \frac{i\hbar}{2m}\,G^{(r)}(z,z';\omega,k_x,k_y), \\
  \lambda_j &=& \frac{2m}{\hbar^2}\,\mu_j, \\
  k^2_0 &=& k_x^2+k_y^2-\frac{2m\omega}{\hbar}.
\end{eqnarray}
To find the reduced Green's function $g^{(r)}(z,z';\omega,k_x,k_y)$, let us rewrite Eq. (\ref{RGF}) as
\begin{equation}\label{2}
 \bigg[\frac{d^2}{dz^2}-k_0^2\bigg]\,g^{(r)}(z,z';\omega,k_x,k_y)=-\delta(z-z')+\sum_{j=1}^n\,\lambda_j \delta(z-a_i)\,g^{(r)}(z,z';\omega,k_x,k_y).
\end{equation}
The free space Green's function of Eq. (\ref{2}) satisfies
\begin{equation}\label{3}
  \bigg[\frac{d^2}{dz^2}-k_0^2\bigg]\,g^{(r)}_0(z,z';k_o)=-\delta(z-z'),
\end{equation}
with some specified boundary conditions on infinite, semi-infinite or finite intervals, for example we may assume $\lim_{z\rightarrow\pm\infty}g^{(r)}_0(z,z';k_0)=0$.  By making use of the Fourier transform, the free-space Green's function can be easily found as
\begin{equation}\label{4}
 g^{(r)}_0 (z,z';k_0)=\frac{e^{-k_0 |z-z'|}}{2k_0}.
\end{equation}
Having the free-space Green's function and using the properties of the delta-function, we can solve Eq. (\ref{2}) as
\begin{eqnarray}\label{5}
 g^{(r)}(z,z';\omega,k_x,k_y) &=& \int dz''\,g^{(r)}_0(z,z'';k_0)\,(\delta(z''-z')-\sum_i \lambda_i \,\delta(z''-a_i)\,g^{(r)}(z'',z';k_0)),\nonumber\\
 &=& g^{(r)}_0(z,z';k_0)-\sum_i \lambda_i \,g^{(r)}_0(z,a_i;k_0)\,g^{(r)}(a_i,z';k_0).
\end{eqnarray}
If we set $z=a_j$ in Eq. (\ref{5}), then by rearranging the terms we have
\begin{equation}\label{6}
\sum_{j=1}^n\,\bigg[\delta_{ij}+\lambda_i\,g^{(r)}_0 (a_j,a_i;k_0)\bigg]\,g^{(r)}(a_i,z';\omega,k_x,k_y)=g^{(r)}_0 (a_j,z';k_0).
\end{equation}
To solve Eq. (\ref{6}), let us define the matrix
\begin{equation}\label{LandC}
\Lambda^{(r)}_{ij}=\delta_{ij}+\lambda_j\,g^{(r)}_0 (a_i,a_j;k_0),
\end{equation}
which is a symmetric matrix due to the symmetry $g^{(r)}_0 (a_j,a_i;k_0)=g^{(r)}_0 (a_i,a_j;k_0)$. Equation (\ref{6}) is a system of $n$ linear equations in $n$ unknowns with the solution
\begin{equation}\label{7}
 g^{(r)}(a_i,z';\omega,k_x,k_y)= \sum_j (\Lambda^{(r)})^{-1}_{ij}\,g^{(r)}_0 (a_j,z';k_0),
\end{equation}
where $(\Lambda^{(r)})^{-1}_{ij}$ are $i,j$-components of the inverse of the matrix $\Lambda^{(r)}$. By inserting Eq. (\ref{7}) into Eq. (\ref{5}) we finally find
\begin{equation}\label{8}
 g^{(r)}(z,z';\omega,k_x,k_y)=g^{(r)}_0 (z,z';k_0)-\sum_{i,j=1}^n \,g^{(r)}_0 (z,a_i;k_0)\lambda_i\,(\Lambda^{(r)})^{-1}_{ij}\,g^{(r)}_0 (a_j,z';k_0).
\end{equation}
Therefore, we only need to compute the inverse of the algebraic matrix $\Lambda^{(r)}$. Let us find the strong coupling limit where the coupling constants tend to infinity which is equivalent to the presence of impenetrable walls. For this purpose we can define $n$ independent unit vectors $\{|e_i\ra\}_{i=1}^n$ spanning a $n$-dimensional vector space. In this space, if we can define the matrices $\boldsymbol{\lambda}=diag(\lambda_1,\lambda_2,\cdots,\lambda_n)$, $ \la e_i|\boldsymbol{\Lambda}^{(r)}|e_j\ra=\Lambda^{(r)}_{ij}$ and $\la e_i|\mathbf{g}^{(r)}_0 |e_j\ra=g^{(r)}_0 (a_i,a _j;k_0)$, then
\begin{equation}\label{9}
\lim_{\{\lambda_k\rightarrow\infty\}} \lambda_i \,(\Lambda^{(r)})^{-1}_{ij}=\lim_{\{\lambda_k\rightarrow\infty\}} \la e_i|\boldsymbol{\lambda}\,\frac{\mathbb{I}}{\mathbb{I}+\boldsymbol{\lambda}\,\mathbf{g}^{(r)}_0}|e_j\ra=\la e_i|(\mathbf{g}^{(r)}_{0})^{-1}|e_j\ra=(g^{(r)}_{0})^{-1}_{ij},
\end{equation}
therefore, in the limiting case $(\lambda_k\rightarrow\infty,\,k=1,\cdots,n)$, using Eqs. (\ref{8},\ref{9}) we have
\begin{equation}\label{10}
  g^{(r)}(z,z';\omega,k_x,k_y)=g^{(r)}_0 (z,z';k_0)-\sum_{i,j=1}^N \,g^{(r)}_0 (z,a_i;k_0)\,(g^{(r)}_{0})^{-1}_{ij}\,g^{(r)}_0 (a_j,z';k_0).
\end{equation}
The poles of the frequency-space Green's function determines the energy spectrum of the constrained system. From Eq. (\ref{10}) it is clear that these poles are in fact the roots of the characteristic equation which is defined as the determinant of the matrix $(g^{(r)}_{0})_{ij}$ appearing in the inverse matrix $(g^{(r)}_{0})^{-1}$. Therefore the characteristic equation is defined by $f(\omega)=\det[g^{(r)}_{0}]$.

\theoremstyle{definition}
\newtheorem{exmp}{Example}
\begin{exmp}
If we set $n=1$, then from (\ref{4}) and (\ref{LandC}) we have
\begin{eqnarray}\label{e1}
&& \lambda(z)=\delta(x-a_1),\\
&& \Lambda_{11}=1+\frac{\lambda}{2k_0}\rightarrow\Lambda^{-1}_{11}=\frac{1}{1+\frac{\lambda}{2k_0}},
\end{eqnarray}
Therefore,
\begin{equation}\label{g1}
  g^{(r)}(z,z';\omega,k_x,k_y)=\frac{e^{-k_0 |z-z'|}}{2k_0}-\frac{\lambda}{1+\frac{\lambda}{2k_0}}\frac{e^{-k_0 |z-a_1|}}{2k_0}\frac{e^{-k_0 |z'-a_1|}}{2k_0}.
\end{equation}
If we set $n=2$, then
\begin{eqnarray}
&& \lambda(z)= \lambda_1\delta(x-a_1)+\lambda_2\delta(x-a_2),\\
&& \Lambda =\left(
            \begin{array}{cc}
              1+\frac{\lambda_1}{2k_0} & \lambda_2\,\frac{e^{-k_0 |a_1-a_2|}}{2k_0} \\
             \lambda_1\,\frac{e^{-k_0 |a_1-a_2|}}{2k_0} & 1+\frac{\lambda_2}{2k_0} \\
            \end{array}
          \right),
\end{eqnarray}
with the inverse
\begin{equation}\label{e2}
  \Lambda^{-1}=\frac{1}{\Delta}\left(
            \begin{array}{cc}
              1+\frac{\lambda_2}{2k_0} & -\lambda_2\,\frac{e^{-k_0 |a_1-a_2|}}{2k_0} \\
             -\lambda_1\,\frac{e^{-k_0 |a_1-a_2|}}{2k_0} & 1+\frac{\lambda_1}{2k_0} \\
            \end{array}
          \right),
\end{equation}
where
\begin{equation}\label{det}
 \Delta=(1+\frac{\lambda_1}{2k_0})(1+\frac{\lambda_2}{2k_0})-\lambda_1\lambda_2\,\frac{e^{-2k_0 |a_1-a_2|}}{4k^2_0},
\end{equation}
is the determinant of $\Lambda^{(r)}$. Therefore,
\begin{eqnarray}
  g^{(r)}(z,z';\omega,k_x,k_y)=g^{(r)}_0 (z-z';k_0) &-& \frac{\lambda_1}{\Delta}(1+\frac{\lambda_1}{2k_0})g^{(r)}_0 (z-a_1;k_0)g^{(r)}_0 (z'-a_1;k_0)
  \nonumber \\
   &+& \frac{\lambda_1 \lambda_2}{\Delta}(\frac{e^{-k_0 |a_1-a_2|}}{2k_0})\,g^{(r)}_0 (z-a_1;k_0)\,g^{(r)}_0 (z'-a_2;k_0)\nonumber \\
   &+& \frac{\lambda_2 \lambda_1}{\Delta}(\frac{e^{-k_0 |a_1-a_2|}}{2k_0})\,g^{(r)}_0 (z-a_2;k_0)\,g^{(r)}_0 (z'-a_1;k_0)\nonumber \\
   &-& \frac{\lambda_2}{\Delta}(1+\frac{\lambda_2}{2k_0})\,g^{(r)}_0 (z-a_2;k_0)\,g^{(r)}_0 (z'-a_2;k_0).
\end{eqnarray}
In the strong coupling limit $(\lambda_1,\,\lambda_2\rightarrow\infty)$, we have
\begin{equation}
  \lambda_i\,(\Lambda^{(r)})^{-1}_{ij}\rightarrow \frac{2k_0}{1-e^{-2k_0 a}}\,\left(
                                           \begin{array}{cc}
                                             1 & -e^{-k_0 \,a} \\
                                             -e^{-k_0 \,a} & 1 \\
                                           \end{array}
                                         \right),
\end{equation}
therefore,
\begin{eqnarray}\label{Recn2}
 g^{(r)}(z,z';\omega,k_x,k_y) = \frac{e^{-k_0 |z-z'|}}{2k_0} &-& \frac{2k_0}{1-e^{-2k_0 a}}\bigg[\frac{e^{-k_0 |z-a_1|}}{2k_0}\frac{e^{-k_0 |a_1-z'|}}{2k_0}\nonumber\\
 &-& \frac{e^{-k_0 |z-a_1|}}{2k_0}\frac{e^{-k_0 |a_2-z'|}}{2k_0} e^{-k_0 a}\nonumber\\
 &-& \frac{e^{-k_0 |z-a_2|}}{2k_0}\frac{e^{-k_0 |a_1-z'|}}{2k_0} e^{-k_0 a}\nonumber\\
 &+& \frac{e^{-k_0 |z-a_2|}}{2k_0}\frac{e^{-k_0 |a_2-z'|}}{2k_0} \bigg],
\end{eqnarray}
where $a=|a_1-a_2|$, is defined for notational simplicity. The characteristic equation in this case is given by
\begin{equation}\label{ch1}
  f(k_0)=\det[g_0^{(r)}]=\det\left(
               \begin{array}{cc}
                 \frac{1}{2k_0} & \frac{e^{-k_0 a}}{2k_0} \\
                 \frac{e^{-k_0 a}}{2k_0} & \frac{1}{2k_0} \\
               \end{array}
             \right)=\frac{1-e^{-2k_0 a}}{4k_0^2}.
\end{equation}
\end{exmp}

\subsection{Cylindrical Geometry}

In cylindrical geometry the walls are located at the positions $\rho=b_1,\,\rho=b_2,\cdots, \rho=b_n$ and the potential is defined by $V(\rho)=\sum\limits_{j=1}^n \nu_j \,\delta(z-b_j)$, where $\nu_j$ are coupling constants. In cylindrical geometry using the azimuthal symmetry we can expand the frequency-space propagator as
\begin{equation}\label{SFC}
  \tilde{K}(\mathbf{x},\mathbf{x}',\omega)=\frac{1}{2\pi}\sum\limits_{m=-\infty}^{\infty}\int \frac{d k_z}{2\pi}\,e^{ik_z (z-z')}e^{im(\varphi-\varphi')}
  G^{(c)}(\rho,\rho';\omega,k_z,m),
\end{equation}
by inserting Eq. (\ref{SFC}) into Eq. (\ref{FS}) we find for the reduced Green's function
\begin{equation}\label{CGF}
 \bigg[\frac{d^2}{d \rho^2}+\frac{1}{\rho}\frac{d}{d\rho}-\frac{m^2}{\rho^2}-k_0^2-\sum\limits_{j=1}^n \lambda_j \,\delta(z-b_j)\bigg]\,g^{(c)}(z,z';\omega,k_z,m)=-\frac{\delta(\rho-\rho')}{\rho},
\end{equation}
where
\begin{eqnarray}
  g^{(c)}(\rho,\rho';\omega,k_z,m) &=& \frac{i\hbar}{2m}\,G^{(c)}(\rho,\rho';\omega,k_z,m), \\
  \lambda_j &=& \frac{2m}{\hbar^2}\,\nu_j, \\
  k^2_0 &=& k^2_z-\frac{2m\omega}{\hbar}.
\end{eqnarray}
In the free space $(\lambda_1=\lambda_2=\cdots=\lambda_n=0)$, and the free-space Green's function is given in terms of the modified Bessel functions as
\begin{equation}\label{c2}
  g^{(c)}_0 (\rho,\rho',k_0)=I_m(k_0 \rho_{<})K_m (k_0 \rho_{>}),
\end{equation}
where $\rho_{<}=\mbox{Min}\{\rho,\rho'\}$ and $\rho_{>}=\mbox{Max}\{\rho,\rho'\}$.
Following the same steps as the previous section, the reduced Green's function in cylindrical geometry is found as
\begin{equation}\label{c3}
 g^{(c)}(\rho,\rho';\omega,k_z,m)=g^{(c)}_0 (\rho,\rho',k_0)-\sum_{i,j=1}^{n}\, g^{(c)}_0 (\rho,b_i,k_0)\,b_i\,\lambda_i\,\Lambda^{-1}_{c,ij}\,g^{(c)}_0 (b_j,\rho',k_0),
\end{equation}
where
\begin{equation}\label{c4}
 \Lambda^{(c)}_{ij}=\delta_{ij}+b_i \lambda_i g^{(c)}_0 (b_j,b_i,k_0).
\end{equation}
In the limiting case $(\lambda_k\rightarrow\infty,\,k=1,\cdots, n)$, we have
\begin{equation}\label{c5}
 g^{(c)}(\rho,\rho';\omega,k_z,m)=g^{(c)}_0 (r,r',k_0)-\sum_{i,j=1}^{n}\, g^{(c)}_0 (r,b_i,k_0)\,(g^{(c)}_0)^{-1}_{ij}\,g^{(c)}_0 (b_j,r',k_0).
\end{equation}
\begin{exmp}
Setting $n=1$, we have
\begin{equation}\label{ec1}
  \Lambda^{(c)}_{11}=1+b_1\,\lambda_1\,I_m (k_0 b_1)K_m (k_0 b_1)\rightarrow(\Lambda^{(c)})^{-1}_{11}=\frac{1}{1+b_1\,\lambda_1\,I_m (k_0 b_1)K_m (k_0 b_1)},
\end{equation}
therefore,
\begin{equation}\label{ec2}
  g^{(c)}(\rho,\rho';\omega,k_z,m)=g^{(c)}_0 (\rho,\rho',k_0)-\frac{b_1 \lambda_1 g^{(c)}_0 (\rho,b_1,k_0) g^{(c)}_0 (\rho',b_1,k_0)}{1+b_1\,
  \lambda_1\,I_m (k_0 b_1)K_m (k_0 b_1)}.
\end{equation}
Setting $n=2$,
\begin{equation}\label{ec3}
  \Lambda^{(c)}=\left(
            \begin{array}{cc}
              1+a_1 \lambda_1 g^{(c)}_0 (b_1,b_1) & b_1 \lambda_1 g^{(c)}_0 (b_1,b_2) \\
             b_2 \lambda_2 g^{(c)}_0 (b_2,b_1) & 1+b_2 \lambda_2 g^{(c)}_0 (b_2,b_2) \\
            \end{array}
          \right),
\end{equation}
and the inverse is
\begin{equation}\label{e2}
  (\Lambda^{(c)})^{-1}=\frac{1}{\Delta'}\left(
            \begin{array}{cc}
             1+b_2 \lambda_2 g^{(c)}_0 (b_2,b_2) & -b_1 \lambda_1 g^{(c)}_0 (b_1,b_2) \\
             -b_2 \lambda_2 g^{(c)}_0 (b_2,b_1) & 1+b_1 \lambda_1 g^{(c)}_0 (b_1,b_1) \\
            \end{array}
          \right),
\end{equation}
where
\begin{equation}\label{det}
 \Delta'=(1+b_1 \lambda_1 g_0 (b_1,b_1))(1+b_2 \lambda_2 g_0 (b_2,b_2))-b_1 b_2 \lambda_1 \lambda_2 g_0 (b_1,b_2)g_0 (b_2,b_1),
\end{equation}
is the determinant of the matrix $\Lambda^{(c)}$. Now the reduced Green's function can be determined in a closed form using Eq. (\ref{c3}). The characteristic equation for $n=2$ is given by
\begin{equation}\label{ch2}
  f(\omega)=\det[g_0^{(c)}]=I_m(k_0 b_1)K_m (k_0 b_1)\,I_m(k_0 b_2)K_m (k_0 b_2)-(I_m(k_0 b_1)K_m (k_0 b_2))^2,\,\,\,\,(b_1<b_2).
\end{equation}
\end{exmp}

\subsection{Spherical Geometry}

In spherical geometry the walls are located at the positions $r=c_1,\,r=c_2,\cdots, r=c_n$ and the potential is defined by $V(r)=\sum\limits_{j=1}^n \eta_j \,\delta(r-c_j)$, where $\eta_j$'s are coupling constants. In the spherical geometry we can expand the frequency-space propagator as
\begin{equation}\label{SFS}
  \tilde{K}(\mathbf{x},\mathbf{x}',\omega)=\sum_{l=-\infty}^{\infty}\sum_{m=-l}^l G^{l}(r,r';\omega)\,Y_{lm}(\theta,\varphi)Y_{lm}^*(\theta',\varphi'),
\end{equation}
and by inserting Eq. (\ref{SFS}) into Eq. (\ref{FS}) we find
\begin{equation}\label{SGF}
  \bigg[\frac{d^2}{d r^2}+\frac{2}{r}\frac{d}{dr}-\frac{l(l+1)}{r^2}-\sum\limits_{j=1}^n\lambda_j\,\delta(r-c_j)-k_0^2\bigg]\,g^l(r,r';\omega_0)=
  -\frac{\delta(r-r')}{r^2},
\end{equation}
where
\begin{eqnarray}
  g^l(r,r';\omega) &=& \frac{i\hbar}{2m}\,G^l(r,r';\omega), \\
  \lambda_j &=& \frac{2m}{\hbar^2}\,\eta_j, \\
  k^2_0 &=& -\frac{2m\omega}{\hbar},
\end{eqnarray}
In the absence of the potential $(\lambda_1=\lambda_2=\cdots=\lambda_n=0)$, the free-space reduced Green's function is given in terms of the modified spherical Bessel functions as
\begin{equation}\label{spfG}
  g^{l}_{0} (r,r',k_0)=-\mbox{i}_l(kr_{<})\, \mbox{k}_l (k r_{>}),
\end{equation}
where $r_{<}=\mbox{Min}\{r,r'\}$ and $r_{>}=\mbox{Max}\{r,r'\}$. In this case the reduced Green's function can be written as
\begin{equation}\label{SPG}
 g^{l} (r,r',k)=g^{l}_{0}(r,r',k_0)-\sum_{i,j=1}^{N}\, g^{l}_{0}(r,c_i,k_0)\,c^2_i\,\lambda_i\,(\Lambda^{(s)})^{-1}_{ij}\,g^{l}_{0}(c_j,r',k_0),
\end{equation}
where
\begin{equation}
  \Lambda^{(s)}_{ij}=\delta_{ij}+c_i^2 \lambda_i \,g^{l}_{0}(c_j,c_i,k_0).
\end{equation}
In the limiting case $(\lambda_k\rightarrow\infty,\,k=1,\cdots,n)$, we have
\begin{equation}\label{spg1}
   g^{l} (r,r',k)=g^{l}_{0}(r,r',k_0)-\sum_{i,j=1}^{n}\, g^{l}_{0} (r,c_i,k_0)\,(g^{l}_0)^{-1}_{ij}\,g^{l}_{0} (c_j,r',k_0).
\end{equation}
and the characteristic equation for $n=2$ is given by
\begin{equation}\label{ch3}
  f(\omega)=\det[g_0^l]=\mbox{i}_l(k c_1)\, \mbox{k}_l (k c_1)\,\mbox{i}_l(k c_2)\, \mbox{k}_l (k c_2)-(\mbox{i}_l(k c_1)\, \mbox{k}_l (k c_2))^2,\,\,\,(c_1<c_2).
\end{equation}
Note that in strong coupling limit we only need to find the inverse of an algebraic matrix defined by the free-space reduced Green's function calculated at the boundary points and the characteristic equation is defined by the determinant of this matrix.

\subsection{Arbitrary operator}

A generalization is straightforward in all these geometries. For example in rectangular geometry we can consider the equation
\begin{equation}\label{g1}
\bigg[\hat{o}-\sum_{i=1}^{N}\lambda_i\,\delta(z-a_i)\bigg]\,g(z,z')=-\delta(z-z'),
\end{equation}
where $\hat{o}$ is an arbitrary linear Hermitian operator. The free-space reduced Green's function satisfies $\hat{o}\, g_0 (z,z')=-\delta(z-z')$ and we assume that $\forall i,\,g_0 (a_i,a_i)\neq \infty$. In higher dimensions where $g_0 (\mathbf{r},\mathbf{r};\omega)$ does not exist a regularization technique may be applied \cite{4}. Following a process similar to the previous sections we will find
\begin{equation}\label{g2}
g(z,z')=g_0 (z,z')-\sum_{i,j=1}^n g_0 (z,a_i)\,\lambda_i\,\Lambda^{-1}_{ij}\,g_0 (a_j,z'),
\end{equation}
where $\Lambda_{ij}=\delta_{ij}+\lambda_j \,g_0 (a_i,a_j)$. In the limiting case $(\lambda_k\rightarrow\infty,\,k=1,\cdots,n)$, we find for the reduced Green's function
\begin{equation}\label{g3}
g(z,z')=g_0 (z,z')-\sum_{i,j=1}^N g_0 (z,a_i)\,(g_0)^{-1}_{ij}\,g_0 (a_j,z'),
\end{equation}
and the characteristic equation is defined by $f(\omega)=\det[g_0 (a_i,a_j)]$.

\section{Confined harmonic oscillator}

 Many attempts have been done for finding the energy eigenvalues of a constrained quantum harmonic oscillator (see \cite{Campoy},\cite{Aguilera}, and references therein). In most of those works, authors tried to solve the problem numerically. In this section, a closed form expression for the characteristic equation of the constrained oscillator is given. The roots of the characteristic equation can be obtained easily to an arbitrary order of approximation. Here the first six eigenvalues of the constrained oscillator are calculated and compared with the numerical data reported in \cite{Aguilera}.

 The finite-box potential can be written as $V(z)=\mu\delta(z)+\mu\delta(z-a)$, and we consider the limit $\mu\rightarrow\infty$ at the end of calculations in order to have the infinite-box potential. The propagator of the confined harmonic oscillator in frequency-space satisfies
 \begin{equation}\label{Har1}
  \bigg[-\frac{\hbar^2}{2m}\frac{d^2}{dz^2}+\haf m\omega_0^2 (z-\frac{a}{2})^2+\mu\delta(z)+\mu\delta(z-a)\bigg]\,G(z,z';\omega)=-i\hbar\,\delta(z-z'),
 \end{equation}
 or equivalently
 \begin{equation}\label{Har2}
  \bigg[\frac{d^2}{dz^2}-\frac{m^2\omega_0^2}{\hbar^2}(z-\frac{a}{2})^2-\lambda\,\delta(z)-\lambda\,\delta(z-a)\bigg]\,g^{os}(z,z';\omega)=-\delta(z-z'),
 \end{equation}
 where $g^{os}(z,z';\omega)=\frac{i\hbar}{2m}\,G(z,z';\omega)$ and $\lambda=2m\mu/\hbar^2$. In this case the operator $\hat{o}$ is defined by
 \begin{equation}\label{Har3}
   \hat{o}=\frac{d^2}{dz^2}-\frac{m^2\omega_0^2}{\hbar^2}(z-\frac{a}{2})^2.
 \end{equation}
Now setting $n=2$ in Eq. (\ref{g3}), we find the Green's function as
\begin{equation}\label{ex51}
  g^{os}(z,z':\omega)=g^{os}_0 (z,z':\omega)-\sum_{i,j=1}^2 g^{os}_0 (z,a_i;\omega)\,(g^{os}_0)^{-1}_{ij}\,g^{os}_0 (a_j,z';\omega).
\end{equation}
The Green's function of the unconstrained harmonic oscillator is given by \cite{Bakh}
\begin{equation}\label{bakhg}
  g^{os}_0 (z,z':\omega)=\sqrt{\frac{m}{\pi\hbar\omega_0}}\,\Gamma(-v)D_v (-y_{<})D_v(y_{>}),
\end{equation}
where $D_v(-y)\,(D_v(y))$ are parabolic cylinder functions \cite{dlmf}, $y_{<}$ and $y_{>}$ are the lesser and greater of $y=\sqrt{2m\omega_0/\hbar}\,(z-a/2)$ and $y=\sqrt{2m\omega_0/\hbar}\,(z'-a/2)$, or equivalently $z$ and $z'$. The dimensionless parameter $v$ is defined by $v=\frac{E}{\hbar\omega_0}-\haf$. The inverse matrix in Eq. (\ref{ex51}) is given by
\begin{equation}\label{inv}
 (g^{os}_0)^{-1}=\frac{1}{\Delta}\left(
                     \begin{array}{cc}
                      g^{os}_0 (a,a;\omega) & -g^{os}_0 (0,a;\omega) \\
                     -g^{os}_0 (a,0;\omega) & g^{os}_0 (0,0;\omega) \\
                     \end{array}
                   \right),
\end{equation}
where the determinant which contains all information about the poles of the Green's function is defined by
\begin{equation}\label{delta}
  \Delta=g^{os}_0 (0,0;\omega)g^{os}_0 (a,a;\omega)-g^{os}_0 (0,a;\omega)g^{os}_0 (a,0;\omega).
\end{equation}
By inserting Eq. (\ref{bakhg}) into Eq. (\ref{delta}), we find the characteristic equation
\begin{equation}\label{chara}
  \Delta(v)=\frac{m}{\pi\hbar\omega_0}\,\Gamma^2(-v)\,D^2_v(\sqrt{2m\omega_0/\hbar}\,a/2)\,
   \big[D^2_v(-\sqrt{2m\omega_0/\hbar}\,a/2)-D^2_v(\sqrt{2m\omega_0/\hbar}\,a/2)\Big].
\end{equation}
The energy eigenvalues satisfy the equation $\Delta(v)=0$. To find the approximate locations of the roots we can equivalently plot the norm function $|\Delta(v)|$. The result is depicted in Fig.1, showing the locations of the first six eigenvalues. To find the locations of the roots more exactly one can restrict the range of the plotting to the values suggested from the Fig.1, for example in Fig.2 the approximate value $v_0=4.45$ is obtained for the fist root when the range of plotting is restricted to the interval $(4.3, 4.8)$. Using simple algorithms of numerical analysis one can find the roots up to an arbitrary order of approximation.
\newpage
\begin{figure}[h!]
   \includegraphics[scale=.7]{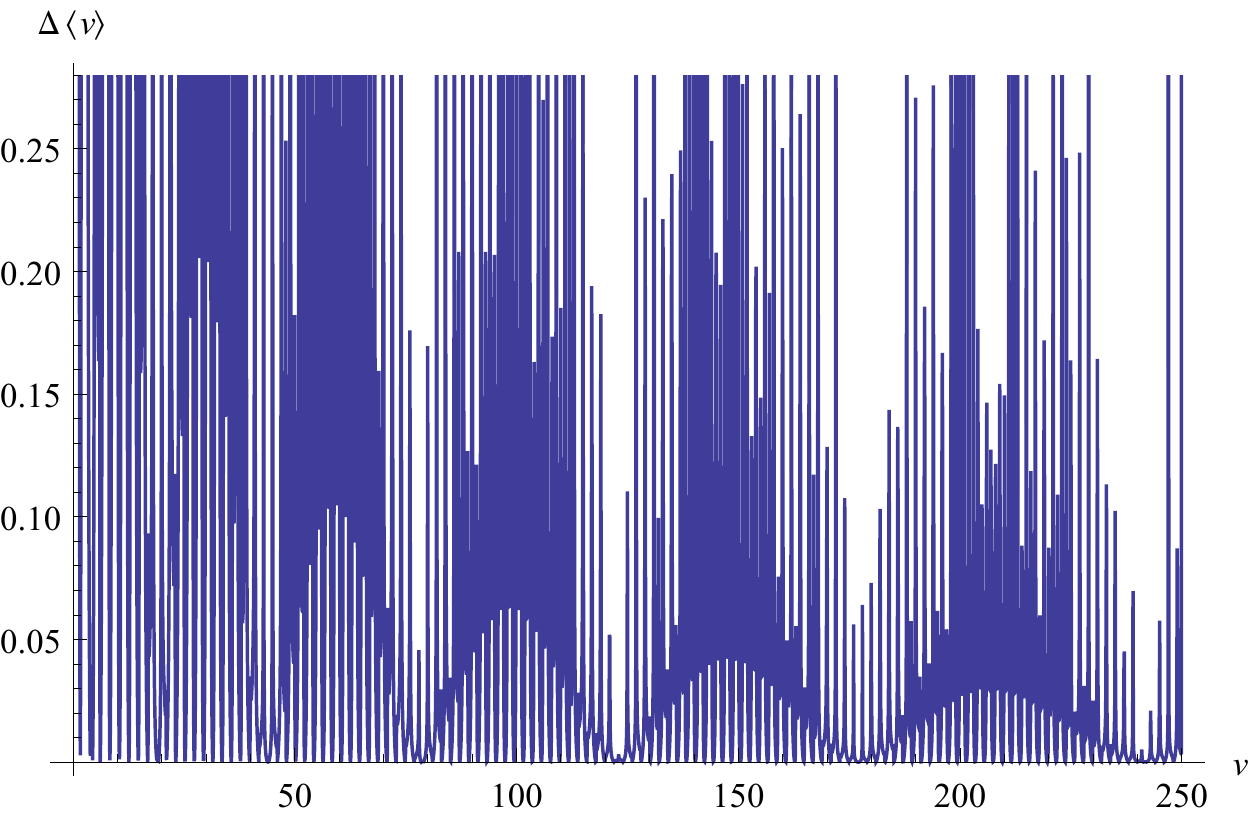}\\
  \caption{The plot of $|f(v)|$ verses $v$. The function $f(v)$ makes fast oscillations around a blurry curve that touches the
  $v$-axis at the approximate points $v_0=4.45,\,v_1=19.27,\,v_2=43.95,\,v_3=78.49,\,v_4=122.91,\,v_5=177.19 $. The approximate eigenvalues are given in Table.1 and compared with the values reported in \cite{Aguilera}. }\label{cap}
\end{figure}
\begin{figure}[h!]
   \includegraphics[scale=.6]{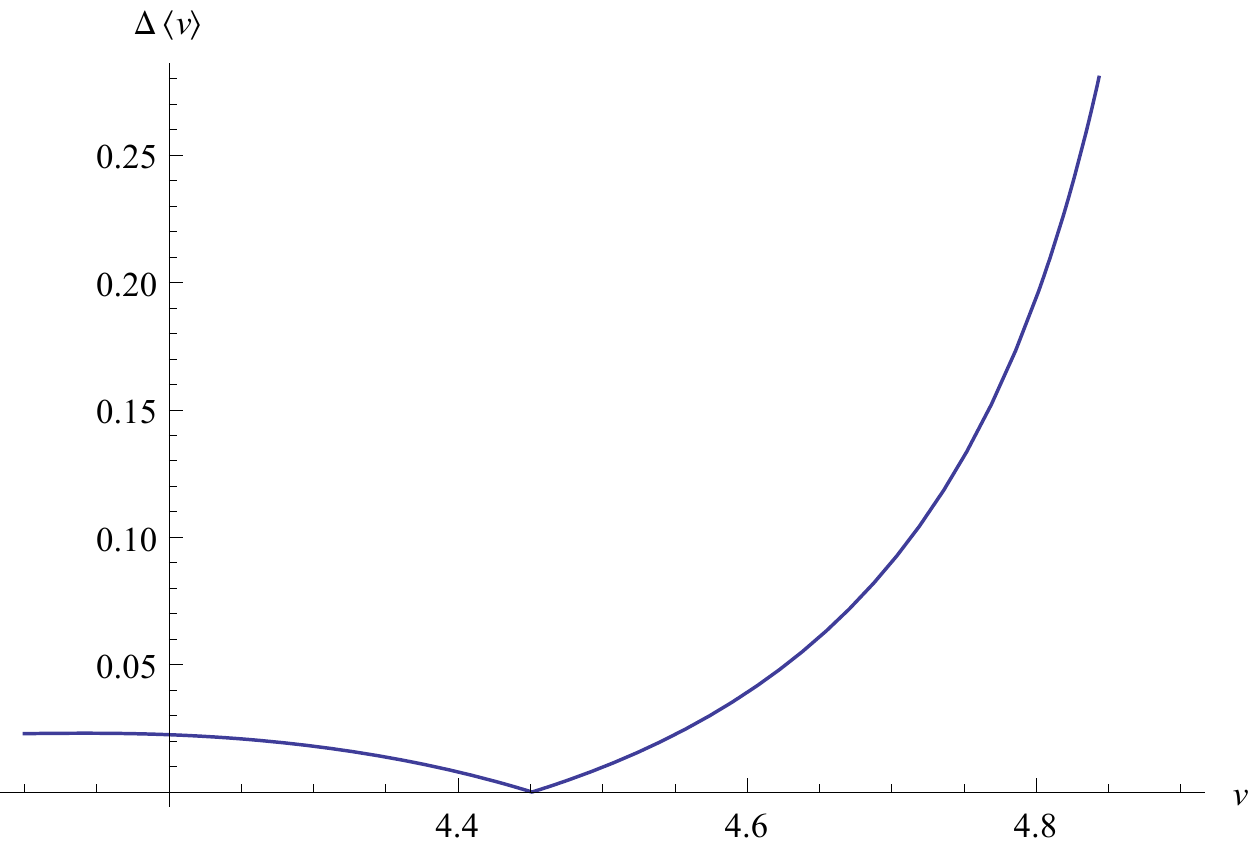}\\
  \caption{The plot of $|f(v)|$ when $v$ is restricted to the interval (4.3, 4.8), showing the approximate location of a root around 4.45.}\label{cap}
\end{figure}
\newpage
\begin{table}[h!]
\begin{tabular}{lll}
\toprule
Energy & \hspace{0.5cm}[14] & \hspace{0.5cm}Ours \\
\colrule
$E_0$ & \hspace{0.5cm}4.951  & \hspace{0.5cm}4.95\\
\colrule
$E_1$ & \hspace{0.5cm}19.774 & \hspace{0.5cm}19.77\\
\colrule
$E_2$ & \hspace{0.5cm}44.452 & \hspace{0.5cm}44.45\\
\colrule
$E_3$ & \hspace{0.5cm}78.996 & \hspace{0.5cm}78.99\\
\colrule
$E_4$ & \hspace{0.5cm}123.410 & \hspace{0.5cm}123.41\\
\colrule
$E_5$ & \hspace{0.5cm}177.693 & \hspace{0.5cm}177.69\\
\botrule
\end{tabular}
\caption{The first six energy eigenvalues calculated from $\Delta(v)=0$ in units of $\hbar\omega_0$. The roots are compared with those reported in \cite{Aguilera}, for the case where the box size $a$ is equal to the oscillator characteristic length $\sqrt{\hbar/m\omega_0}$, (unit length).}
\label{tab1}
\end{table}

\section{conclusion}

Closed analytic representations of the reduced Green's functions in the presence of a chain of $\delta$-potentials are given in the rectangular, cylindrical and spherical coordinate systems. From these reduced Green's functions corresponding quantum propagator or kernel can be obtained easily. The method is easy and efficient compared to the usual method of piecewise solution of a differential equation and applying the continuity and discontinuity conditions at each boundary point. The strong coupling limit $(\{\lambda_k\}\rightarrow\infty)$ of the reduced Green's functions are obtained and discussed. In strong coupling limit the characteristic equation is defined by $\det[g_0 (a_i,a_j;\omega)]=0$, and its roots determines the energy eigenvalues. For a quantum harmonic oscillator confined to an infinite-box, exact characteristic equation is derived and its first six roots are calculated approximately and compared with the existing numerical data.

\section{Suggested problem}

Problem. In this article we assumed that the free-space reduced Green's function fulfills $\forall i,\,g_0 (a_i,a_i)\neq \infty$. In higher dimensions where $g_0 (\mathbf{r},\mathbf{r};\omega)$ does not exist a regularization technique is needed. How we can generalize the method to higher dimensions? the interested reader may consult \cite{4}.

\end{document}